\documentclass[twocolumn,prb,showpacs,amsmath,amssymb]{revtex4-1}
\usepackage[dvipdfmx]{graphicx}
\usepackage{bm}
\begin{document}
\title{Coflow turbulence of superfluid ${}^4$He in a square channel: Vortices trapped on a cylindrical attractor}
\author{Shinichi Ikawa$^1$}
\author{Makoto Tsubota$^{1,2}$}
\affiliation{$^1$Department of Physics, Osaka City University, 3-3-138 Sugimoto, Sumiyoshi-Ku, Osaka 558-8585, Japan}
\affiliation{$^2$The OCU Advanced Research Institute for Natural Science and Technology (OCARINA), Osaka City University, 3-3-138 Sugimoto, Sumiyoshi-Ku, Osaka 558-8585, Japan}
\date{\today}

\begin{abstract}
We perform a numerical simulation of the dynamics of quantized vortices produced by  coflow in a square channel using the vortex filament model. 
Unlike the situation in thermal counterflow, where the superfluid velocity $\bm{v}_s$ and normal fluid velocity $\bm{v}_n$ flow in opposite directions, in coflow, $\bm{v}_s$ and $\bm{v}_n$ flow in the same direction. 
Quantum turbulence in thermal counterflow has been long studied theoretically and experimentally, and its various features have been revealed.
In recent years, an experiment on quantum turbulence in coflow has been performed to observe different features of thermal counterflow. 
By supposing that $\bm{v}_s$ is uniform and  $\bm{v}_n$ takes the Hagen-Poiseuille profile, which is different from the experiment where  $\bm{v}_n$ is thought to be turbulent, we calculate the coflow turbulence.
Vortices preferentially accumulate on the surface of a cylinder for $\bm{v}_s \simeq \bm{v}_n$ by mutual friction; namely, the coflow turbulence has an attractor.
How strongly the vortices are attracted depends on the temperature and velocity.
The length of the vortices increases as the vortices protruding from the cylindrical attractor continue to wrap around it.
As the vortices become dense on the attractor, they spread toward its interior by their repulsive interaction.
Then, the superfluid velocity profile induced by the vortices gradually mimics the normal-fluid velocity profile.
This is an indication of velocity matching, which is an important feature of coflow turbulence.
\end{abstract}

\maketitle

\section{Introduction \label{introduction}}
Quantum turbulence is one of the most important issues in low-temperature physics and has been studied for more than half a century {\cite{tsubota}. 
Superfluid ${}^4$He is the most typical system in which quantum turbulence is realized.
Quantum turbulence can be created in pipe flow of superfluid ${}^4$He in several ways.
The typical case is thermal counterflow, in which the superfluid velocity and normal fluid velocity have opposite directions.
Many experimental and theoretical studies of thermal counterflow have been accumulated {\cite{tough}.
There is another case of pipe flow, namely, coflow, where the superfluid velocity and normal-fluid velocity have the same direction, but it has seldom been studied.
A recent experiment on coflow observed features that differed from those of thermal counterflow {\cite{varga}.
The motivation of the present paper is to find numerically some remarkable behavior of vortices characteristic of coflow. 

The hydrodynamics in superfluid ${}^4$He is described mainly by a two-fluid model and quantized vortex {\cite{donnelly}.
The liquid state of ${}^4$He exists in two phases: a high-temperature phase called He I and a low-temperature phase called He II.
According to the two-fluid model, He II below a $\lambda$ point of 2.17 K is regarded as an intimate mixture of a viscous normal fluid and an inviscid superfluid. 
The density and velocity of the normal fluid are denoted by $\rho_{\rm n}$ and $\bm{v}_{\rm n}$, whereas those of the superfluid are denoted by $\rho_{\rm s}$ and $\bm{v}_{\rm s}$, respectively. 
The total density $\rho=\rho_{\rm n}+\rho_{\rm s}$ is almost independent of temperature below the $\lambda$ point. 
However, the relative proportions of the normal fluid and the superfluid, $\rho_{\rm n}/\rho$ and $\rho_{\rm s}/\rho$, depend strongly on the temperature. 
The velocity fields are independent unless quantized vortices are relevant.

The idea of quantized circulation, which was considered by Onsager {\cite{onsager} and confirmed by Vinen {\cite{vinen61}, mentions that in He II, the circulation of superfluid flow is quantized by the quantized circulation $\kappa=h/m_4$, where $h$ is Planck's constant, and $m_4$ is the mass of a ${}^4$He atom. 
The elementary excitations forming the normal fluid are strongly scattered as vortex lines appear at finite temperatures.
Thus, if there is a relative velocity between the normal fluid and the quantized vortices, a frictional force called a mutual friction force works between them {\cite{gorter}.
The mutual friction term is taken into account in the dynamics of the two fluids, so the two fluids become coupled.

We briefly review the history of quantum turbulence to clarify the context of our research.
Quantum turbulence manifests itself as a tangle of vortex lines {\cite{feynman} and can be generated in many ways.
Thermal counterflow is the first type of turbulent flow that was studied in detail, in a series of pioneering papers by Vinen; it is explained by a two-fluid model {\cite{vinen57a,vinen57b,vinen57c,vinen57d}.
A channel is prepared with one end connected to a He II bath and the other end closed.
A heat current is applied to the closed end of the channel; then the normal fluid will flow from the warm side to the cool side, while the superfluid will flow in the opposite direction to conserve the total mass:
\begin{equation}
  \int ( \rho_{\rm n} {\bm v}_{\rm n} + \rho_{\rm s} {\bm v}_{\rm s} ) dS = {\bm 0} ,
  \label{mass}
\end{equation}
where the integral is performed over the cross section of the channel.
Thus, a relative velocity $v_{\rm ns} = \overline{ |{\bm v}_{\rm n}-{\bm v}_{\rm s}| }$ occurs between the two fluids, where the overline denotes the spatial average over the channel cross section.
When the counterflow velocity exceeds a critical value, a self-sustaining tangle of quantized vortices appears, forming superfluid turbulence.
Measurements \cite{vinen57c} show that the vortex line density (VLD) $L$ follows the relation
\begin{equation}
  L^{1/2} = \gamma (v_{\rm ns} - v_{\rm 0}),
  \label{vinen}
\end{equation}
where $\gamma$ is a parameter depending on the temperature, and $v_{\rm 0}$ is a critical velocity that determines whether the vortex tangle remains or vanishes.
In typical experiments, $v_{\rm 0}$ is much smaller than $v_{\rm ns}$, so $v_{\rm 0}$ is usually negligible.

A scheme for understanding quantum turbulence in terms of the vortex dynamics was considered by Vinen.
By assuming homogeneous turbulence, he estimated the vortex growth using a dimensional analysis and modeled the decay process phenomenologically \cite{vinen57c}.
He showed that the dynamics of the vortex tangle is described by  Vinen's equation,
\begin{equation}
  \frac{dL}{dt} = \chi_1 v_{\rm ns} L^{3/2} - \chi_2 L^{2} ,
\label{vinen2}
\end{equation}
where $\chi_1$ and $\chi_2$ are temperature-dependent parameters.
If the vortex tangle is in a steady state, $dL/dt =0$, resulting in Eq. (\ref{vinen}).

Schwarz investigated counterflow quantum turbulence using the vortex filament model and dynamical scaling \cite{schwarz88}.
The observable quantities obtained in his calculation agree with the experimental results for vortex tangles in the steady state.
However, his simulation could sustain the steady vortex tangle only through an artificial mixing procedure.
Adachi {\it et al.} argued that this is the result of using the localized induction approximation (LIA), in which the interaction between vortices is neglected \cite{adachi}.
By performing the full Biot--Savart calculation, they overcame the difficulty and successfully obtained a steady state consistent with the experimental results.

The above studies were performed by assuming homogeneous turbulence.
However, the visualization experiment performed by Marakov {\it et al.} shows that the normal-fluid profile in a pipe is actually nonuniform because of the boundary of the channel {\cite{marakov}.
By taking account of the boundary's effect on quantum turbulence, Baggaley {\it et al.} introduced the Poiseuille profile as the normal-fluid velocity in two parallel plates {\cite{baggaley13,baggaley14}.
This simulation found that vortices are distributed inhomogeneously and that the physical quantities have a spatial dependence. 
Yui {\it et al.} introduced the Hagen--Poiseuille profile as the normal-fluid velocity in a square channel and found inhomogeneous turbulence with a superfluid boundary layer {\cite{yui}.

In contrast to the research activity related to thermal counterflow, coflow turbulence induced by a mass flow has not attracted much attention.
In recent years, Varga {\it et al.} performed an experiment on coflow driven mechanically by a bellows through a square channel {\cite{varga}.
This experiment observed features different from those of thermal counterflow.
For example, the VLD is proportional to the 3/2 power of the velocity and is independent of temperature.
These results indicate that the VLD does not obey Eq. (\ref{vinen}) and that the dynamics of coflow cannot be explained by Eq. (\ref{vinen2}).
Numerical studies of coflow could find other characteristic features that differ from those of thermal counterflow turbulence, in addition to the above observations.
Note that the normal fluid flow as well as the superfluid flow is thought to be turbulent in the experiment {\cite{skrbek}.

Our simulation \cite{movie} shows that vortices are localized on the surface of a cylinder for $\bm{v}_s \simeq \bm{v}_n$.
That is, our system has an attractor for vortices.
Then, the superfluid velocity profile induced by the vortices gradually mimics the normal-fluid velocity profile.
This is nothing but velocity matching.
Velocity matching is an important feature of coflow and appears experimentally and theoretically in various situations.
Experiments in coflow were performed in many ways in addition to the method used by Varga {\it et al.} \cite{varga}.
For example, coflow is induced by spinning disks or propellers, towing a grid or sphere, and rotating cylinders (Taylor--Couette flow).
These experiments \cite{walstrom, borner, smith, maurer, smith93} showed that at low velocity, the two fluids are independent; however, at high velocity, they appear to be coupled and behave as a single Navier--Stokes fluid with density $\rho = \rho_{\rm n} + \rho_{\rm s}$ and the viscosity of the normal fluid.
One interpretation is that the superfluid velocity field is driven to match the normal-fluid velocity field through the mutual friction.
A simulation of coflow performed by Samuels showed velocity matching \cite{samuels}.
However, this study used some approximations and simplifications.
We obtain an indication of velocity matching without these approximations and simplifications.

In this paper, we perform a numerical analysis of coflow in a square channel.
Although the normal flow in the above experiment seems turbulent, in this paper we assume that the normal flow is laminar in order to focus on the low-velocity condition.
The Reynolds number $Re$ of the normal component calculated in our simulation is $350 \leq Re \leq 1100$. It is smaller than the critical Reynolds number, which characterizes the transition from laminar to turbulent flow and is about 2000 in typical cases {\cite{davidson}. 
Therefore, the normal-fluid profile might be regarded as laminar.
We use the Hagen--Poiseuille profile for the normal-fluid profile because our simulation includes the effect of boundaries in a square channel. 

The contents of this paper are as follows. 
Section ~\ref{formulation} clarifies the formulation of the model and the equation of motion. 
In Sec. ~\ref{coflow}, we show the characteristics of coflow turbulence by using the full Biot--Savart law and some physical parameters.
In Sec. ~\ref{vel}, we investigate velocity matching in which the superfluid velocity profile matches the normal-fluid velocity.
Section ~\ref{conclusions} presents the conclusion and describes future work. 

\section{Formulation \label{formulation}}
In this section, we describe the formulation and numerical analysis of a vortex filament model {\cite{schwarz85}. 
A quantized vortex is defined by a filament passing through the fluid and has a definite direction corresponding to its vorticity. 
This approximation is suitable in He II because the core size of the quantized vortex is much smaller than any other characteristic length scale. 
Except in the core region, a superfluid velocity field has a classically well-defined meaning and can be described by ideal fluid dynamics. 
Then, the velocity produced at a point $\bm{r}$ by a filament is given by the Biot--Savart expression 
\begin{equation}
  {\bm v} _{\rm{s},\omega} ({\bm r})=
    \frac{\kappa}{4 \pi} \int _{\cal L}
      \frac{ ({\bm s} _{1} - {\bm r}) \times d {\bm s} _{1} }{ |{\bm s} _{1} - {\bm r} |^{3} }
      .
\end{equation}
The filament is represented in parametric form as ${\bm s}={\bm s}(\xi,t)$, where ${\bm s}_1$ refers to a point on the filament, and the integration is performed along the filament. 
Attempting to calculate the velocity ${\bm v} _{\rm{s},\omega}$ at a point $\bm{r} = \bm{s}$ on the filament makes the integral diverge as $\bm{s}_1 \rightarrow \bm{s}$. 
To avoid this, we divide the velocity $\dot{\bm{s}}$ of the vortex filament at the point $\bm{s}$ into local and nonlocal terms:
\begin{equation}
\dot{\bm s}=
\beta {\bm s}' \times \bm{s}'' +
    \frac{\kappa}{4 \pi} \int _{\cal L}^{'}
      \frac{ ({\bm s} _{1} - {\bm s}) \times d {\bm s} _{1} }{ |{\bm s} _{1} - {\bm s} |^{3} }
      .
\label{full_vortex_velocity}
\end{equation}
Here, the prime denotes derivatives of $\bm{s}$ with respect to the coordinate $\xi$ along the filament, and $\beta$ takes a value proportional to the quantum circulation. 
The first term refers to the localized induction field arising from a curved line element acting on itself. 
The second term represents the nonlocal field obtained by performing the Biot--Savart integral along the rest of the filament and all other filaments in the system. 
When a solid boundary exists, the velocity in the direction perpendicular to the wall must vanish, so the image vortex is described by reflecting the vortex filament into the surface and reversing its direction. 
The velocity produced by an image vortex is denoted by $\bm{v}_{\rm{s,b}}$. 
If a velocity applied by an external field exists, it is denoted by $\bm{v}_{\rm{s,a}}$. 
At zero temperature, the vortex filament moves with the total superfluid velocity 
\begin{equation}
\bm v_{\rm s}=
\bm{v}_{\rm{s},\omega}
      +\bm{v}_{\rm{s,b}}
      +\bm{v}_{\rm{s,a}}
      .
\label{superfluid_velocity}
\end{equation}
At finite temperatures, the mutual friction due to the interaction between the vortex core and the normal fluid is taken into account. 
The velocity of a point $\bm{s}$ is then given by
\begin{equation}
\dot{\bm{s}}=
\bm{v}_{\rm s}
+\alpha\bm{s}'\times(\bm{v}_{\rm n}-\bm{v}_{\rm s})
-\alpha'\bm{s}'\times[\bm{s}'\times(\bm{v}_{\rm n}-\bm{v}_{\rm s})]
,
\label{vortex_motion}
\end{equation}
where $\alpha$ and $\alpha'$ are the temperature-dependent coefficients. 

By using the LIA, in which the second term in Eq. (\ref{full_vortex_velocity}) is neglected, we can understand the role of mutual friction.
By neglecting the term with $\alpha '$, we obtain 
\begin{equation}
\dot{\bm{s}}=
\beta {\bm s}' \times \bm{s}''
+\bm{v}_{\rm{s,a}}
+\alpha\bm{s}'\times(\bm{v}_{\rm n}-\bm{v}_{\rm{s,a}}-\beta {\bm s}' \times \bm{s}'')
.
\label{LIA}
\end{equation}
The third term causes a curved vortex to balloon outward or collapse inward. 
As discussed by Schwarz \cite{schwarz85},  when the relative velocity $\bm{v}_{\rm ns}=\bm{v}_{\rm n}-\bm{v}_{\rm s,a}$ flows against $\beta {\bm s}' \times \bm{s}''$, the mutual friction always shrinks the curved vortex locally. 
On the other hand, $\bm{v}_{\rm ns}$ flowing along $\beta {\bm s}' \times \bm{s}''$ yields a critical radius of curvature $R_{\rm c}$. 
When the local radius of curvature $R$ at a point on the vortex is smaller than $R_{\rm c}$, the curved vortex shrinks locally, whereas the curved vortex balloons out when $R>R_{\rm c}$. 

In this study, we prescribe uniform flow for $\bm{v}_{\rm s,a}$ and the Hagen--Poiseuille profile  $u_{\rm p}$ for $\bm{v}_{\rm n}$. 
When the normal fluid flows along the ${\it x}$ direction, the ${\it x}$ component of $\bm{v}_{\rm n}$ is represented by 
\begin{equation}
\begin{split}
  u_{\rm p}(y,z) =& u_0 \sum _{m=1,3,5,\cdots} ^{\infty} (-1)^{\frac{m-1}{2}}
  \\
  &
  \times
  \left[
    1 - \frac{ \cosh(m \pi z / 2 a) }{ \cosh(m \pi b / 2 a) }
  \right]
  \frac{ \cos(m \pi y / 2 a) }{ m^3 }
  ,
\end{split}
\label{poi}
\end{equation}
where $u_0$ is a normalization factor, and $a$ and $b$ are the half-channel widths along the $y$ and $z$ axes, respectively {\cite{poiseuille}. 

To characterize the dynamics of vortices, we introduce some statistical values.
The VLD is given by 
\begin{equation}
L = \frac{1}{\Omega}\int_{\cal L}d\xi,
\end{equation}
where the integral is performed along all vortices in the sample volume $\Omega$. 
The anisotropy parameter \cite{schwarz88} is defined as
\begin{equation}
  I = \frac{1}{\Omega L} \int _{\cal L} [1-({\bm s}' \cdot \hat{\bm r}_{\mathrm p})^2 ] d\xi,
\end{equation}
where $\hat{\bm r}_{\mathrm p}$ represents the unit vector parallel to the flow direction.
When the vortex tangle is completely isotropic, $I = 2/3$.
When the tangle consists entirely of curves lying in the plane normal to the flow direction, $I=1$.

The simulation is performed under the following conditions. 
We discretize the vortex lines into a number of points held at a minimum spatial resolution of $\Delta\xi=8.0\times10^{-4}$ cm.
Integration in time is achieved using a fourth-order Runge--Kutta scheme with a time resolution of $\Delta t=1.0\times10^{-4}$ s. 
The computational box is $0.1\times0.1\times0.1$ ${\rm cm}^3$ in size, and $a$ and $b$ in Eq.  (\ref{poi}) are 0.05 cm. 
We regard the velocity condition of coflow as ${\bm{v}}_{\rm {s,a}}=\bar{\bm{v}}_{\rm n}$, where $\bar{\bm{v}}_{\rm n}$ is the spatially averaged normal fluid velocity. 
Periodic boundary conditions are used along the flow direction $x$, whereas solid boundary conditions are applied to the channel walls. 
The effects of reconnection are artificially applied whenever two vortices approach more closely than $\Delta\xi$. 
The initial state consists of eight randomly oriented vortex rings of radius 0.023 cm [Fig. \ref{snapshot2}(a)]. 

\section{Numerical simulation of coflow turbulence \label{coflow}}
In this section, we present the results of our simulation under the Hagen--Poiseuille flow expressed by Eq. (\ref{poi}).
First, we show that in the coflow dynamics, vortices are attracted to a cylindrical region and exhibit polarization. 
Second, we discuss why and how the vortices have this attractor.
Finally, by introducing a statistical value to classify the vortex configuration, we realize that the vortex configuration of coflow has parameter dependence.

\subsection{Dynamics \label{dynamics}}
The coflow dynamics differs in some ways from the thermal counterflow dynamics.
To characterize the coflow dynamics, we show the statistical values and typical snapshot of the dynamics in Figs. \ref{statistical} and \ref{snapshot}, respectively. 
As shown in Fig. \ref{statistical}[a], the VLD just increases, not reaching any steady state. 
However, as we explain later, the VLD of coflow should have a steady state like that of thermal counterflow.
On the other hand, the anisotropic parameter $I$ shown in Fig. \ref{statistical}[b] becomes steady at $I \simeq$ 0.85.
This shows that the vortices in coflow are strongly anisotropic because $I$ is larger than the isotropic value, $I = 2/3$.
The snapshots in Figs. \ref{snapshot}[a] and \ref{snapshot}[b] show that most vortices are localized in a cylindrical region and polarized along the flow direction \cite{movie}; that is, the vortices appear to be trapped by an attractor {\cite{edward}.
The properties of the attractor are discussed in Sec. \ref{attractor}.
\begin{figure}
  \includegraphics[width=0.4\textwidth]{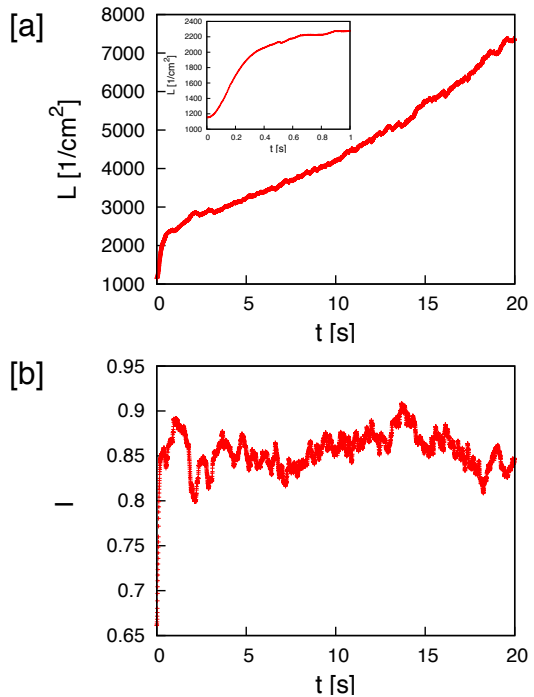}
  \caption
  { Time development of the VLD [a] and the anisotropic parameter [b] for $T$ = 1.85 K and $\bar{\bm{v}_{\rm n}}(=\bar{\bm{v}_{\rm s}}) = 1.2$ cm/s.
Inset in [a] shows the time development of the VLD in the very early stage, 0 s $\leq t \leq$ 1.0 s.}
  \label{statistical}
\end{figure}
\begin{figure}
  \includegraphics[width=0.5\textwidth]{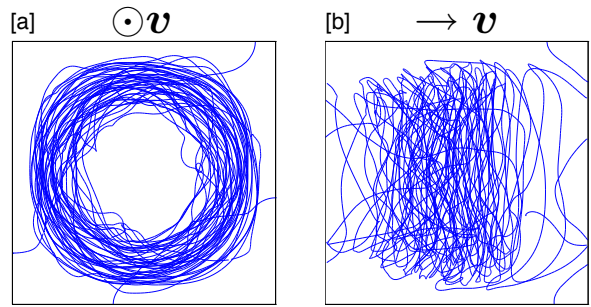}
\caption{Snapshots of vortices at $t=$ 20 s in Fig. \ref{statistical}: [a] viewed along the flow direction and [b] viewed from the side of the flow direction.}
\label{snapshot}
\end{figure}

The time development of the vortices can be characterized by two stages. 
As shown in Fig. \ref{statistical}[a], the VLD increases rapidly in the first stage, 0 s $\leq t <$ 0.4 s, and then it increases slowly in the second stage, 0.4 s $\leq t$; the mechanism by which the vortices multiply is different in the two stages.
In the first stage, shown in Fig. \ref{snapshot2}, the initial vortices expand and make many reconnections, which create numerous small vortices.
All these vortices repeat the process until they are trapped by the attractor.
Figure \ref{snapshot} shows a typical snapshot of the second stage.
Most vortices are localized in the attractor, and a few vortices protrude from the attractor toward the walls.
In the second stage, there are mainly two mechanisms that slowly increase the VLD.
The first mechanism is ``wrapping the attractor.''
The protruding vortices rotate around the attractor because of the mutual friction.
Then, the vortex edges on the attractor leave traces on it, increasing the vortex length on the attractor.
The second mechanism is ``spreading inside.''
When the vortices become dense on the attractor, their repulsive interaction makes them spread toward the interior and increases the VLD.
The two mechanisms increase the VLD continuously.
Because reconnections occur much less frequently than in the first stage, the VLD increases much more slowly than in the first stage.
Because the two mechanisms approximately maintain anisotropy, the anisotropy parameter takes a steady value, although the VLD continues to increase.
\begin{figure}[h]
  \includegraphics[width=0.4\textwidth]{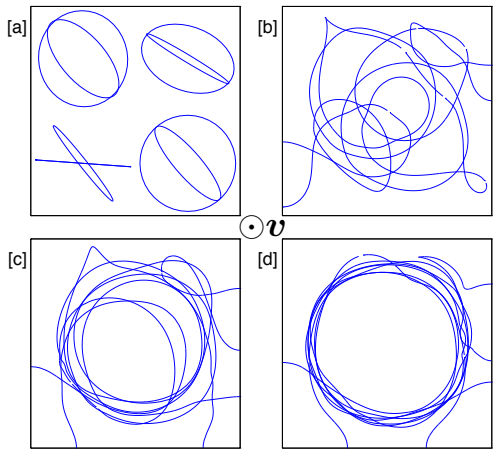}
\caption{Snapshots of the time development of the vortex tangle in the first stage viewed along the flow direction ($T$ = 1.85 K, $\bar{\bm{v}_n}$ = 1.2 cm/s): [a] $t = 0$ s, [b] $t = 0.15$ s, [c] $t = 0.25 $ s, [d] $t = 0.4$ s.}
\label{snapshot2}
\end{figure}

\subsection{Attractor for vortices  \label{attractor} }
In the previous section, we showed that vortices are localized in a cylindrical region.
The appearance of the attractor is an important characteristic of the present system.
This section discusses why the attractor appears and what determines the topological region. 

We consider the situation in a circular pipe where the normal-fluid profile is Poiseuille flow to understand analytically why and how the vortices are localized.
For the sake of simplicity, we assume that a vortex ring is placed with cylindrical symmetry, namely, that a vortex ring moves along the central axis of a circular pipe.
The vortex ring is represented in cylindrical coordinates as $\bm{s} = \bm{s}(R,\theta,z)$.
Then, Eq. (\ref{LIA}) is reduced to
\begin{equation}
\frac{dz}{dt} =
\frac{\beta}{R}+\bm{v}_{\rm s,a}
\label{LIA_z}
\end{equation}
and
\begin{equation}
\frac{dR}{dt} =
\alpha \left(v_{\rm n} - v_{\rm s,a} - \frac{\beta}{R}\right)
.
\label{LIA_R}
\end{equation}
Here, the dynamics of $\theta$ is irrelevant because this system is axisymmetric.
We focus only on the dynamics of $R$ because the motion of $z$ is irrelevant to the localization.
The normal fluid profile is prescribed to be the Poiseuille profile
\begin{equation}
v_{\rm n}(R) = V_{\rm n} \left[ 1- \left(\frac{R}{D}\right)^2 \right]
,
\label{poiseuille_profile}
\end{equation}
where $D$ is the radius of the pipe, and $V_{\rm n}$ is the maximum value of $v_{\rm n}$.
By inserting Eq. (\ref{poiseuille_profile}) into Eq. (\ref{LIA_R}), we obtain
\begin{equation}
\frac{dR}{dt} =
\alpha \left(v_{\rm s,a}  \left[ 1- 2\left(\frac{R}{D}\right)^2 \right] - \frac{\beta}{R}\right)
,
\label{LIA_poi}
\end{equation}
where we use the condition $\bar{\bm{v}}_{\rm n}={\bm{v}}_{\rm {s,a}}$.
The stationary state of $R$ is given by 
\begin{equation}
R\left[ 1- 2\left(\frac{R}{D}\right)^2 \right] =\frac{\beta}{ v_{\rm s,a}}.
\label{LIA_poi2}
\end{equation}
If the right-hand side of Eq. (\ref{LIA_poi2}) is negligible, we have only one solution for $R > 0$, namely, $R=D/\sqrt{2}$.
When the right-hand side of Eq. (\ref{LIA_poi2}) is not negligible, we have two solutions for $R > 0$.
However, the smaller solution does not correspond to the localized position because this state is unstable.
Vortices are localized at the larger solution because this state is stable.
Consequently, the cylindrical region of the radius corresponding to the larger solution acts as an attracter where the mutual friction vanishes for vortex rings {\cite{edward}.

However, in a square channel, the geometry of the attractor is modified from a cylindrical region because the Hagen--Poiseuille profile is not axisymmetric.
We consider the region where the mutual friction of Eq.  (\ref{LIA}) vanishes.
 First, if the term $\beta\bm{s}'\times\bm{s}''$ is negligible, the region of the attractor is given by the condition $\bm{v}_{\rm{n}} =\bm{v}_{\rm{s,a}}$ and shown by the light symbols in Fig. \ref{fig:10}. 
However, this region is modified to the result shown by dark symbols by the term $\beta\bm{s}'\times\bm{s}''$; the position around the corners is shifted inward because the radius of curvature is small, and the position around the sides is shifted outward because it is large {\cite{schwarz85}.
In this way, the appearance of the attractor causes the vortex dynamics of coflow to differ from that of nonuniform thermal counterflow \cite{yui} because the coflow has a region where mutual friction does not work.
\begin{figure}[h]
\includegraphics[width=0.5\textwidth]{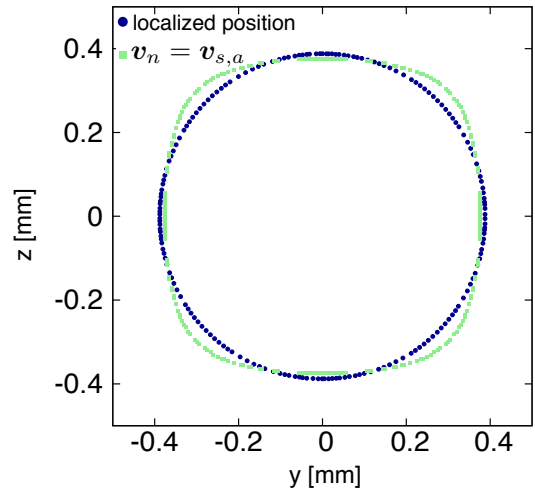}
\caption{Region where mutual friction vanishes: the dark symbols show the region where  $\bm{v}_{\rm{s,a}} =\bm{v}_{\rm{n}}$, and the light symbols represent  $\bm{v}_{\rm{s}} =\bm{v}_{\rm{n}}$.}
\label{fig:10}
\end{figure}
\subsection{Parameter dependence of the vortex configuration}
If we change the temperature $T$ and the spatially averaged velocity $\bar{\bm v} ( = \bar{\bm{v}}_{\rm n}=\bar{\bm{v}}_{\rm {s}} )$, the vortex configuration also changes.
Consequently, the coflow dynamics has a critical velocity $v_{\rm c}$ that depends on the temperature.
When $|\bar{\bm{v}}| < v_{\rm c}$, the vortices are diffusive in all directions without localization, as shown in Fig. \ref{disorder}.
\begin{figure}
\includegraphics[width=0.5\textwidth]{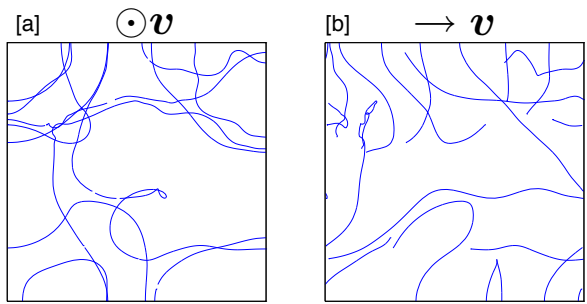}
\caption{Typical snapshots of vortices in the diffusive state: [a] viewed along the flow direction and [b] viewed from the side of the flow direction.}
\label{disorder}
\end{figure}
This diffusive state occurs because the effect of vortex accumulation in the attractor is weak.
In this state, most vortices vanish eventually,  leaving only several vortices in the four corners of the channel.
These vortices rarely disappear even after sufficient time has passed.
When $|\bar{\bm{v}}| \simeq v_{\rm c}$, vortices are localized or diffuse repeatedly.
This state arises from two competing effects.
One is that the mutual friction tends to accumulate vortices in the attractor, as described in Sec. \ref{attractor}.
The other is that the self-induced velocity $\beta\bm{s}'\times\bm{s}''$ preferentially removes vortices from the attractor.
If the effect of the mutual friction is stronger than that of the self-induced velocity, vortices are localized and continue to increase.
Conversely, if not, vortices diffuse and eventually vanish.
In this competing state, we cannot determine easily whether vortices are localized or diffuse, although we may be able to determine this if we continue to calculate for a very long time.

We show the time development of the VLD for three states in Fig. \ref{VLD_phasediagram}[a].
When $|\bar{\bm{v}}| > v_{\rm c}$, which is 0.8 cm/s at $T=1.55$ K, the VLD continues to increase, as shown by the circular symbols.
When $|\bar{\bm{v}}| \simeq v_{\rm c}$, the VLD is statistically steady, as shown by the triangular symbols.
When $|\bar{\bm{v}}| < v_{\rm c}$, the VLD decreases, as shown by the square symbols.
By changing $T$ and $\bar{\bm{v}}$, we classify the time development of the VLD into the above three states and show the phase diagram in Fig. \ref{VLD_phasediagram}[b]: circular symbols denote the localized state described in Sec. \ref{dynamics}, triangular symbols denote the competing state, and square symbols denote the diffusive state.
This shows that the critical velocity $v_{\rm c} (T)$ is a decreasing function.
\begin{figure}
\includegraphics[width=0.4\textwidth]{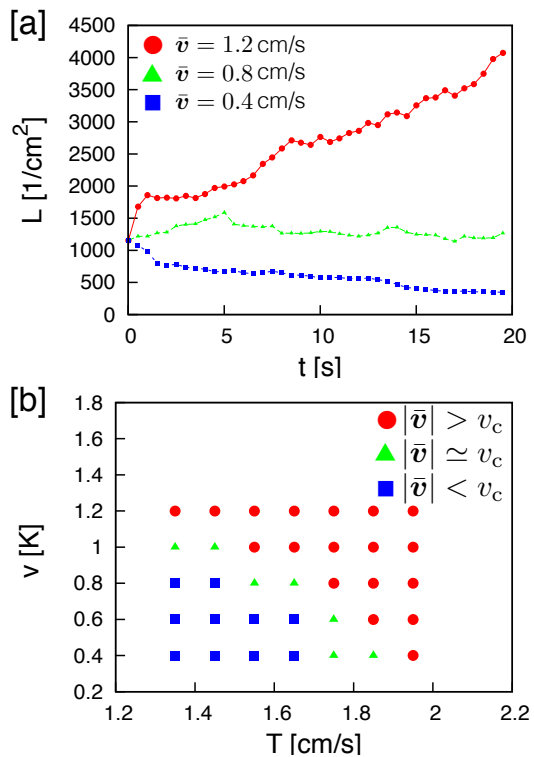}
\caption{[a] Typical development of $L$ with $T=1.55$ K for each state. [b] Phase diagram for the critical velocity.}
\label{VLD_phasediagram}
\end{figure}

We introduce a dimensionless variable $L_{{\rm in}}/L_{{\rm out}}$ to show quantitatively the parameter dependence of the vortex configuration: 
\begin{equation}
\frac{L_{\rm in}}{L_{\rm out}} = \frac{\int_{{\cal L}_{\rm in}}d\xi}{\int_{{\cal L}_{\rm out}}d\xi}.
\end{equation}
Here $L_{\rm in}$ is obtained by integration along all the vortices ${\cal L}_{\rm in}$ in the cylindrical region between the central axis and a radius of $0.045$ cm, and $L_{\rm out}$ is obtained by integration along all the other vortices ${\cal L}_{\rm out}$ in the region between a radius of $0.045$ cm and the wall. 
As the vortices are localized, $L_{\rm in}/L_{{\rm out}}$ increases.
The time development of $L_{{\rm in}}/L_{{\rm out}}$ is shown in Fig. \ref{Lin/Lout}. 
When the time average of $L_{{\rm in}}/L_{{\rm out}}$ is taken, these values increase at higher temperature and larger velocity. 
Consequently, this shows that the vortices tend to be localized at higher temperature and larger velocity, and diffusive for lower temperature and smaller velocity.
\begin{figure}[h]
  \includegraphics[width=0.4\textwidth]{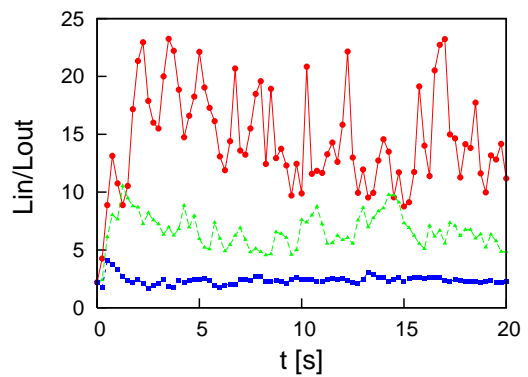}
\caption{Time development of $L_{\rm in}/L_{\rm out}$ for the parameters described in Fig. \ref{VLD_phasediagram}[a].}
\label{Lin/Lout}       
\end{figure}

Mutual friction causes the dependence of $T$ and $\bar{\bm{v}}$ in Figs. \ref{VLD_phasediagram} and \ref{Lin/Lout}.
The mutual friction depends on $T$ and $\bar{\bm{v}}$ because the coefficient $\alpha$ depends on the temperature, and increasing the averaged velocity makes the relative velocity faster everywhere. 
If the temperature is lower and the velocity is smaller, the velocity caused by mutual friction is dominated by the self-induced velocity $\beta\bm{s}'\times\bm{s}''$, and the vortices move almost freely. Consequently, whether the vortices are localized or diffusive depends on the temperature and velocity.

\section{velocity matching \label{vel}}
As noted in Sec. \ref{introduction}, previous experiments \cite{walstrom, borner, smith, maurer, smith93} showed that at low velocity, the two fluids are independent, whereas at high velocity, they appear to be coupled and behave as a single Navier--Stokes fluid with density $\rho = \rho_{\rm n} + \rho_{\rm s}$ and the viscosity of the normal fluid.
One interpretation is that the superfluid velocity field is driven to match the normal-fluid velocity field through the mutual friction.
Velocity matching is one of the most important features of coflow.
Our simulation shows velocity matching from the vortex dynamics, which previous experiments had not addressed.

The pioneer research on velocity matching was done by Samuels \cite{samuels}.
He performed a numerical simulation using the vortex filament model under the condition that the normal fluid has a Poiseuille profile in a circular pipe and $\overline{\bm{v}_{\rm s}} = \overline{\bm{v}_{\rm n}}$, where $\bm{v}_{\rm s}$ is described by Eq. (\ref{superfluid_velocity}).
This research shows that the coflow turbulence has a cylindrical attractor where vortices are localized, namely, attractor, and the superfluid velocity profile gradually mimics the normal-fluid velocity profile.
However, this research used some approximations and simplifications because it was difficult to describe the formula for an image vortex in a circular pipe.
First, to generate vortices regularly in an attractor, the initial arrangement of the vortices is simplified as follows.
A small half ring is placed on the boundary of the pipe.
The radius of the ring is smaller than the distance between the attractor and the boundary.
Once the dynamics starts, the vortex becomes trapped on the attractor by mutual friction and generates another small half ring attached at the boundary by reconnection.
This small half ring also follows the same process, and this event repeats periodically.
This process was followed numerically, neglecting the image vortex.
Second, Samuels approximated the trapped vortices as a group of perfect vortex rings that are polarized along the flow direction.
Then, the image vortex ring for the trapped vortex ring is approximately represented by following form:
\begin{eqnarray}
\label{samuels_image_r}
r_{\rm image} &=& R^2 / r_{\rm ring}, \\
\kappa_{\rm image} &=& -\frac{R}{r_{\rm ring}}\kappa.
\label{samuels_image_kappa}
\end{eqnarray}
Here $r_{\rm image}$ and $\kappa_{\rm image}$ are the radius of curvature and the circulation of the image vortex ring, respectively; $R$ is the pipe radius, and $r_{\rm ring}$ and $\kappa$ are the radius of curvature and the circulation of the vortex, respectively.
Using Eqs. (\ref{samuels_image_r}) and (\ref{samuels_image_kappa}) to describe the image vortex ring is a suitable approximation only when the vortex ring is close to the boundary.
Although Samuels obtained velocity matching by using these processes, this dynamics is not realistic.

We directly obtain an indication of velocity matching from the vortex dynamics without imposing the approximations and simplifications used by Samuels.
The superfluid velocity profile obtained using Eq. (\ref{superfluid_velocity}) in the localized state is shown in Fig. \ref{vs_profile}; the velocity is high in the interior region, where vortices are localized, and low in the exterior region.
\begin{figure}[h]
\includegraphics[width=0.5\textwidth]{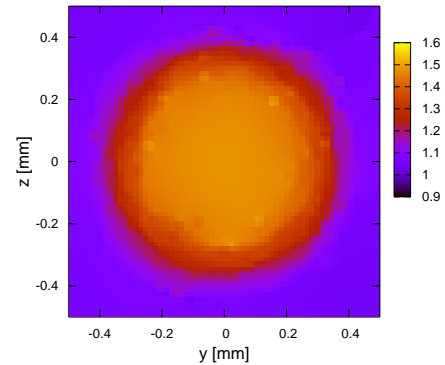}
\caption{Spatial dependence of the superfluid velocity profile on $x=0$ mm at $t=$ 40 s  with $T$ = 1.95 K and $\bar{\bm{v}}=1.0$ cm/s.}
\label{vs_profile}
\end{figure}
Figure \ref{vs_vn} shows the cross section on $y=0$ of Fig. \ref{vs_profile}.
\begin{figure}[h]
\includegraphics[width=0.5\textwidth]{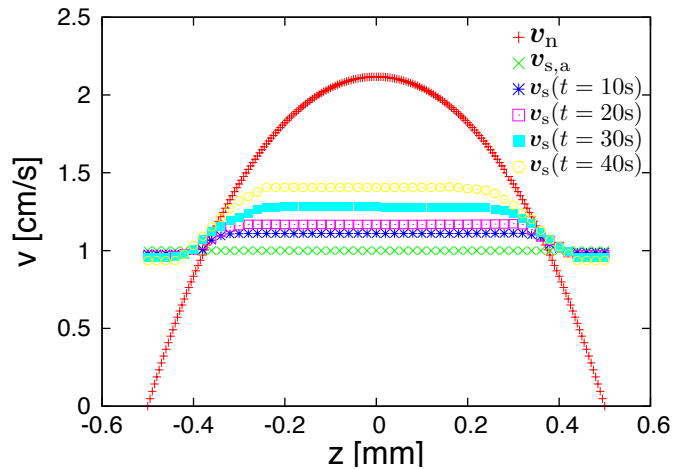}
\caption{Normal fluid velocity profile and time development of superfluid velocity profile.}
\label{vs_vn}
\end{figure}
It shows that the superfluid velocity profile gradually mimics the normal-fluid velocity profile from the attractor toward the central axis of the square channel.
This behavior is induced by ``spreading inside'' as described in Sec. \ref{dynamics}.
The velocity matching might be completed if vortices filled the interior attractor.
We cannot follow the dynamics until the final stage because the vortices are extremely dense.

Another difference between Samuels' work {\cite{samuels} and ours is the way that $\bm{v}_{\rm s,a}$ is handled.
Samuels dynamically adjusted $\bm{v}_{\rm s,a}$ to always satisfy the velocity condition $\overline{\bm{v}_{\rm s}} = \overline{\bm{v}_{\rm n}}$; because $|\overline{\bm{v}_{\rm s,\omega}}|$ increases with time, $|\bm{v}_{\rm s,a}|$ should be reduced.
On the other hand, our study takes  $\overline{\bm{v}_{\rm n}} = {\bm{v}_{\rm s,a}}$ without the ``dynamical adjustment."
Our approach would be acceptable by the following reasons.

 It is difficult and arbitrary to adjust $|\bm{v}_{\rm{s,a}}|$ in the numerical simulation.
Actually we have the same problem in the case of thermal counterflow, where Eq. (\ref{mass}) should be satisfied.
Almost all simulations since Schwarz's pioneering work \cite{schwarz88} fix $\bm{v}_{\rm s,a}$ and $\bm{v}_{\rm n}$, although the development of the vortex tangle increases $|\overline{\bm{v}_{\rm s,\omega}}|$ and eventually the condition Eq. (\ref{mass}) is broken as shown in the recent simulation \cite{yui}.
In order to keep exactly the condition, we have to calculate ${\bm{v}_{\rm s,\omega}}$ by the full Biot-Savart integral at each time step and adjust $\bm{v}_{\rm s,a}$, but it is very difficult numerically.
One way to avoid the difficulty is to make the adjustment not at each step but at each several steps, but this is more or less arbitrary.
We have the same difficulty in the case of coflow.
Samuels did the adjustment only by a too simplified method, not making the full Biot-Savart integral \cite{samuels}.
The numerical simulation taking proper account of the ``dynamical adjustment " would be a future work. 

\section{Conclusions\label{conclusions}}
In this study, we investigated coflow turbulence with  nonuniform flow of a normal fluid using the vortex filament model.
The velocity profile of the normal fluid was prescribed to be the Hagen--Poiseuille profile {\cite{poiseuille}, although in the experiment performed by Varga {\it et al.}, it seems to be turbulent {\cite{varga}.

The most important feature of coflow turbulence is that it has a cylinder in which the normal-fluid velocity equals the superfluid velocity.
Vortices are localized on the surface of the cylinder by mutual friction \cite{movie}; that is, the coflow dynamics has an attractor \cite{edward}.
How strongly the vortices are attracted depends on the temperature and the velocity, because the mutual friction also depends on them.
Thus, a critical velocity appears depending on the temperature.
If the velocity exceeds the critical velocity, the vortices grow.
On the other hand, if it is smaller than the critical velocity, the vortices do not grow.
When the velocity exceeds the critical velocity, the VLD increases as the vortices protruding from the attractor continue to wrap around it.
When the vortices become dense on the attractor, they spread toward the interior of the attractor by their repulsive interaction.
Then, the superfluid velocity profile induced by the vortices gradually mimics the normal-fluid velocity profile.
This is nothing but velocity matching, which is an important feature of coflow turbulence.
Velocity matching in coflow was studied by Samuels \cite{samuels}.
However, his study was done under some approximations and simplifications.
We directly obtained an indication of velocity matching from the vortex dynamics without imposing these approximations and simplifications, although our simulation was not performed to the final state.

In this study, we suppose that the normal-fluid velocity profile is laminar, which is different from the experiment \cite{varga}, where the normal-fluid velocity is thought to be turbulent; therefore, our future work is to perform a calculation under the condition that the normal-fluid velocity is turbulent.
\acknowledgments
We acknowledge L. Skrbek, E. Varga, and S. Babuin for useful discussions.
M. T. was supported by JSPS KAKENHI Grant No. 26400366 and MEXT KAKENHI ``Fluctuation \& Structure,'' Grant No. 26103526.


\end{document}